\documentstyle[prd,aps,psfig]{revtex}
\bibstyle{unsrt}

\tighten
\begin{document}
\draft

\preprint{LAUR-99-6836}
\twocolumn[\hsize\textwidth\columnwidth\hsize\csname 
@twocolumnfalse\endcsname

\title{Properties of the Langevin and Fokker-Planck equations
for scalar fields and their application to the dynamics of 
second order phase transitions}
\author{Lu\'{\i}s M. A. Bettencourt}
\address{Theoretical Division, Los Alamos National Laboratory, Los Alamos
NM 87545}
\date{\today}
\maketitle

\begin{abstract}
I consider several 
Langevin and Fokker-Planck classes of dynamics for scalar field theories
in contact with a thermal bath at temperature $T$.
These models have been applied recently in the numerical description  of 
the dynamics of second order phase transitions and associated topological 
defect formation as well as in other studies of these critical phenomena. 
Closed form solutions of the Fokker-Planck equation
are given for the harmonic potential and a dynamical mean-field 
approximation is developed. These methods allow for an analytical  
discussion of the behavior of the theories in several circumstances 
of interest such as critical slowing down at  
a second order transition and the development  of spinodal 
instabilities. These insights allow for a more detailed understanding
of several numerical studies in the literature.
\end{abstract}

\pacs{PACS Numbers : 05.70.Ca, 11.30.Qc, 02.50.-r, 98.80.Cq \hfill   LAUR-99-6836}

\vskip2pc]

The dynamics of systems undergoing a second order symmetry breaking phase 
transition has been studied recently in several instances \cite{AB,LZ,YZ,ABZ}, 
particularly in association with the theory of topological defect formation 
\cite{Kibble,Zurek} in cosmology \cite{Book} and  in 
experiments in $^3He$ \cite{He3} and $^4He$ \cite{He4}. With experimental 
data in the pipeline for the collisions of heavy ions at RHIC and later 
at CERN the effort to identify phase transitions in nuclear matter is 
also a very active field of research. The Langevin dynamics of effective 
scalar theories (such as the $\sigma$ model) affords one of the few 
quantitative windows into such transitions \cite{BRS}.   

Theoretically this situation has been modeled by the 
out-of equilibrium dynamics of a classical scalar field theories  
(with a given number of flavors) in contact with an external environment
at a temperature T, which may be a function of time. 
The environment is only known statistically and its behavior can
then be described in terms of stochastic fields obeying a 
fluctuation-dissipation relation. The effective evolution 
of the fields is therefore described by a Langevin field equation. 

This stochastic classical description is of course only an approximation
to the full quantum evolution. Recently attempts to treat quantum field 
theories dynamically \cite{Cooper} have been made but the possibilities 
afforded by these approximations thus far do not allow for the correct 
description of thermalization processes \cite{BW} 
(with the exception of Boltzmann transport theory \cite{KB}) and have 
therefore very limited 
applicability to the description of the long time 
dynamics of a system undergoing a symmetry breaking phase transition.

In addition a reasonable case can be made that close to a second order 
transition the theory is effectively at high temperature $T>> m(T)$, where
$m(T)$ is the temperature dependent mass scale, and its infrared 
should behave approximately classically. 
In this regime some quantum effects can also be 
included effectively through the appropriate choice of the coupling 
of the fields to the external environment (the quantum fluctuations 
in this case) and/or their  contributions to the mass and couplings 
of the dynamical fields. This line of considerations \cite{Bodeker} 
has been used 
recently to describe the effective evolution of the long wave-length 
modes of the color fields of non-Abelian gauge theories 
in the high temperature regime, as a stochastic classical field theory.  

In cosmology and high energy physics field theories are relativistic.
This implies that the time evolution of scalar fields 
is governed by a second order derivative in time. 
The canonical Langevin equations  however are usually 
formulated in terms of a first derivative evolution in time. This difference
is unimportant in equilibrium, but modifies real-time properties. In certain 
circumstances, which I discuss in detail below,  the second order evolution 
effectively  leads to a first order (or overdamped) description 
of certain parts of the system. 

The general properties of Langevin field evolutions are that, 
under very minimal conditions, the fields equilibrate for long 
times to the canonical (classical) Boltzmann distribution. 
The dynamics are very rich, realizing and generalizing  
all the near canonical equilibrium properties of the classical 
field evolution, including in particular the features captured by the 
perhaps more familiar time-dependent Ginzburg Landau formulation (TDGL).

In this paper I present a didactic introduction to the underlying 
formulation of a second time derivative Langevin system and its associated 
Fokker-Planck equation. Its general dynamical properties are compared 
to those of other important classes of dynamics such as TDGL and the 
stochastic non-linear Schr\"odinger equation (sNLS).
This is done mostly in section \ref{FP}. (The treatment of the sNLS 
system is given in Appendix A.) In 
section \ref{Sol} I derive analytic solutions of the Fokker-Planck 
equation in the particular case of the harmonic potential. 
These solutions are fairly simple but to the best of my knowledge have not 
been discussed in the literature of stochastic field theories.
A mean-field formulation of the dynamical problem is given in 
section \ref{MF} and it is found that its solution is similar 
in effort to solving the associated Langevin mean-field equation, 
but with the stochastic averages already taken 
into account. In section \ref{Dyn} I discuss applications of the 
solutions of section \ref{Sol} to the understanding of the full non-linear 
dynamics of the system. Although qualitative it will be argued that these 
solutions capture the essentials of the full dynamics in the critical 
domain and permit an understanding of the relevant thermalization times. 
A brief discussion of spinodal instabilities is also included.
Finally in section \ref{Conc} I summarize the conclusions of this paper. 

\section{The Langevin and Fokker-Planck equations for scalar fields}
\label{FP}

We start by formulating the problem. The second order Langevin equation for 
a scalar fields $\phi(x,t)$ with an arbitrary interaction potential 
$V(\phi)$ is given by  
\begin{eqnarray}
\left( \partial_t^2  -\nabla^2 \right) \phi(x) +
 {\delta V(\phi) \over \delta \phi(x)} + \eta \dot \phi(x)= \xi({\bf x},t).
\label{e1}
\end{eqnarray}
The stochastic fields  $\xi(x,t)$ are taken to be Gaussian and white, 
characterized by
\begin{eqnarray}
\langle \xi({\bf x}, t) \rangle = 0,   \quad  
\langle \xi ({\bf x},t) \xi({\bf x'},t') \rangle = \Omega \delta^3 
({\bf x}-{\bf x'})\delta (t-t').
\label{e2}
\end{eqnarray}
Although we have written this equation for a single 
scalar field $\phi(x,t)$ the generalization of Eq.~(\ref{e1}-\ref{e2}) 
and of what follows, to a $O(N)$ symmetric theory, with N the number 
of flavors, is straightforward. 

We can proceed to  reduce the order of this differential 
system, by introducing  field generalized momenta conjugate to $\phi$:  
\begin{eqnarray} 
\partial_t \phi(x) &=& \pi(x), \label{e4a} \\
\partial_t \pi(x) &=& -\eta \pi(x)  + \nabla^2 \phi(x) 
- {\delta V(\phi) 
\over \delta \phi(x)} + \xi(x).
\label{e4b}
\end{eqnarray}

An interesting limit of Eqs.~(\ref{e4a}-\ref{e4b}) arises when 
$\eta$ is large, or more precisely when $\partial_t \pi(x) << \eta \pi(x)$. 
Then we can write
\begin{eqnarray} 
\eta \partial_t \phi(x) &=&  \nabla^2 \phi(x) - {\delta V(\phi) \over 
\delta \phi(x)} + \xi(x).
\label{e5}
\end{eqnarray}
Although this is the more conventional form of the Langevin field equation 
\cite{Zinn,Parisi} I will refer to it as the overdamped limit of 
Eqs.~(\ref{e4a}-\ref{e4b}).

The objective of the Langevin analysis is to measure the 
expectation value (over the stochastic fields) 
of any functional $\rho$ of the fields $\pi(x),\phi(x)$, generated by the 
evolution of Eq~(\ref{e1}-\ref{e2}). 
With the choice Eq.~(\ref{e2}) this average is a Gaussian functional 
integral of the form
\begin{eqnarray}
\int D\xi \rho[\pi_{\xi},\phi_{\xi}] e^{- {1 \over 2 \Omega }
\int d^4 x \xi^2}
\label{e6}
\end{eqnarray}
where the $\xi$ subscripts in the field and its momentum denote their 
functional dependence on the stochastic fields $\xi(x,t)$, 
through the evolution of Eq.~(\ref{e4a}-\ref{e4b}). 
Relation (\ref{e6}) and (\ref{e2}) 
can be generalized to fields $\xi$ with more complicated 
Gaussian distributions.

In the presence of the stochastic fields $\xi(x,t)$ the initial conditions 
for the fields $\pi(x), \ \phi(x)$ may only be know statistically themselves. 
They can then be expressed in terms of a functional probability 
distribution $P[\pi,\phi]$, at the initial time.   

Throughout the evolution we  assume that a time dependent 
probability distribution $P_{\rm FP} [\phi(x),\pi(x),t]$, a  functional 
of the time independent fields and a function of time, exists. This is the 
Fokker-Planck probability 
distribution, which we require, as usual,  to be positive definite and 
normalizable in the sense:
\begin{eqnarray}
{\cal N} \int D\pi D\phi  P_{FP} [\phi,\pi,t] = 1.
\label{e7}
\end{eqnarray}
Note that in general $\cal N$ as well as other expectation values of the 
fields may be formally infinite as the ultraviolet cutoff of the theory is 
taken to zero.

The time dependent expectation value over the 
stochastic fields $\xi$ of any quantity 
$\rho[\phi,\pi]$ is then given by 
\begin{eqnarray}
\langle \rho \rangle(t) = {\cal N} \int D\pi D\phi  P_{FP} [\phi,\pi,t] 
\rho[\phi,\pi].
\label{e8}
\end{eqnarray}

In order to be useful this picture requires the explicit knowledge of 
$P_{FP} [\phi,\pi,t]$.
The Langevin equation generates a time-dependent (Fokker-Planck) 
probability distribution  $P_{FP}[\pi,\phi,t]$, which  we can write formally
\begin{eqnarray}
&& P_{FP}[\pi,\phi,t] = \nonumber \\ 
&& \qquad \langle  
\int d^D x  \delta \left[ \hat \pi(x,t) - \pi(x) \right] 
\delta \left[ \hat \phi (x,t) - \phi(x) \right] \rangle,
\label{e9}
\end{eqnarray}
where the brackets denote, as before, average over the stochastic fields.
The static fields are the arguments of $P_{FP}[\pi,\phi,t]$, whereas 
$\hat \pi(x,t), \ \hat \phi(x,t)$ obey the Langevin 
Eqs. (\ref{e4a}- \ref{e4b}).

The probability $P_{FP}[\pi,\phi,t]$, obeys a functional 
Fokker-Planck evolution equation that can  be computed directly by 
differentiating Eq.~(\ref{e9}). Following 
Zinn-Justin \cite{Zinn} 
\begin{eqnarray}
&& \partial_t P_{FP} = \langle \left[ \int d^D x~ 
\partial_t  \hat \phi (x,t) {\delta \over \delta \hat \phi(x,t)} + 
\partial_t \hat \pi(x,t)   {\delta \over \delta \hat \pi(x,t)} 
\right] \nonumber \\
&&   \qquad \quad \times  \delta \left[ \hat \pi(x,t) - \pi(x) \right] 
\delta \left[ \hat \phi (x,t) - \phi(x) \right] \rangle. 
\label{e10}
\end{eqnarray}
The properties of the $\delta$ allow us to trade the functional derivatives of 
$\hat \pi(x,t), \hat \phi(x,t)$ for derivatives relative to $\pi(x), \phi(x)$.
These can be taken out of the stochastic average $\langle ... \rangle$. The only remaining 
term inside this average results from the appearance of the stochastic fields $\xi$ in the 
equation of motion for $\hat \pi (x,t)$. This term is  of the form 
\begin{eqnarray}
\langle \int d^D x \xi (x,t)  \delta \left[ \hat \pi(x,t) - \pi(x) \right] \rangle.
\end{eqnarray}
The Gaussianity of the stochastic fields enforces the identity \cite{Zinn}
\begin{eqnarray}
&& \langle \xi(x,t)  \delta \left[ \hat \pi(x,t) - \pi(x) \right] \rangle
= \Omega \langle {\delta  \over \delta \xi (x,t)} \delta \left[ \hat \pi(x,t) 
- \pi(x) \right] \rangle \nonumber \\
&& = \Omega \langle {\delta \hat \pi(x,t) \over 
\delta \xi(x,t)}
{\delta \over  \hat \pi(x,t)} \delta \left[ \hat \pi(x,t) 
- \pi(x) \right] \rangle.
\end{eqnarray}
Once again we can trade the functional derivative relative 
to $\hat \pi (x,t)$ for another relative to $\pi (x)$. 
The expectation value of ${\delta \hat \pi(x,t) \over 
\delta \xi(x,t)}$ can in turn be obtained from the formal integration of the 
equation of motion by using a regularized delta function in time. We must 
effectively take half of this delta function to obtain
\begin{eqnarray}
{\delta \hat \pi(x,t) \over \delta \xi(x,t)} = {1\over 2}.
\end {eqnarray}
Bringing together all the terms results finally in
\begin{eqnarray}  
\partial_t P_{FP}[\pi,\phi,t]=-{\cal H}_{\rm FP} P_{FP}[\pi,\phi,t].
\label{e11}
\end{eqnarray}
where
\begin{eqnarray}
{\cal H}_{\rm FP} &=& - {\Omega \over 2} 
{\delta^2 \over \delta \pi^2 } + \pi {\delta \over \delta 
\phi} \nonumber \\
&& \qquad \qquad - {\delta \over \delta \pi} 
\left( \eta \pi - \nabla^2 \phi + 
{\delta V(\phi) \over \delta \phi}\right) 
\label{e12}
\end{eqnarray}

It is useful to think in terms of the functional  operator 
${\cal H}_{\rm FP}$  as the generator of infinitesimal time displacements 
of the probability functional $P_{\rm FP}$.
If, as in most applications, the potential $V(\phi)$ is explicitly time 
independent we can invoke a separation ansatz for $P_{FP}$ such that 
\begin{eqnarray}
P_{FP} [\pi,\phi,t] = {\cal P} [\pi,\phi] T(t)
\label{e13}
\end{eqnarray}   
Thus we  can regard Eq.~(\ref{e11}) as an analog of 
a functional Schr\"odinger equation, in imaginary time. 
Then we can write the time independent and dependent equations 
\begin{eqnarray}
{\cal H}_{\rm FP} {\cal P}_n = E_n {\cal P}_n, \quad \partial_t T(t) = 
- E_n T(t).
\label{e14}
\end{eqnarray}
The functional dependence on the fields is now limited to 
the static probability eigenfunctionals ${\cal P}_n$. The time evolution
of the Fokker-Planck distribution is completely characterized by the spectrum
of eigenvalues  of ${\cal H}_{\rm FP}$, $E_n$.
An orthogonal complete basis of functionals $B_n$ 
can in general be constructed from the 
set ${\cal P}_n$\footnote{The set of functionals  ${\cal P}_n$ will 
only be guaranteed to be an orthogonal basis 
if ${\cal H}_{\rm FP}$ is a Hermitian functional 
operator, which is not true in general. The relation between the 
$B_n$ and $P_n$ is usually very simple involving a power of the canonical
distribution and it follows that $E_n$ are also the eigenvalues of the 
set $B_n$. See {\it eg.} \cite{Risken} for some examples in few degree 
of freedom problems.}
Note that it is not possible in general to separate Eq.~(\ref{e13}) further
into two equations, one in $\pi$ and another in $\phi$ via an 
ansatz like $P_{FP} = R[\pi] S[\phi]$.

On general grounds we expect a steady state corresponding to thermal 
equilibrium to be reached for long times.
Formally, we can then project the evolution of $P_{FP}$ in terms of the 
eigenvalues $E_n$ and functionals $B_n$ as:
\begin{eqnarray}
P_{FP}[\pi,\phi,t] = \sum_{n=0} ^\infty C_n  B_n[\pi,\phi]  
e^{- E_n t}.
\label{e15}
\end{eqnarray}
where the $C_i$'s are the projections of $P_{FP}$ at the initial time onto
the basis of eigenfunctionals $B_n$:
\begin{eqnarray}
C_n = \int D \pi D \phi B_n [\pi,\phi] P_{FP} [ \pi, \phi,t=0],
\label{e16}
\end{eqnarray} 
where both $B_n$ and $P_{FP} [ \pi, \phi,t]$ are taken to be 
normalized.

If equilibrium is approached for long times the corresponding distribution 
must be associated with the zero mode of ${\cal H}_{\rm FP}$, $E_0=0$ i.e. 
it is stable. 
Contributions from eigenstates ${\cal P }_n$, $n \neq 0$,  
vanish exponentially, in a characteristic time 
$t_{\rm eq} \sim E_n^{-1}$, provided that the real part of $E_n$ is 
positive. This follows from the requirement that $V(\phi)$ is bounded
from below. For the $E_n$, solutions of the second time derivative 
eigenproblem there may be in addition an imaginary part. Thus the excited 
states decay away in time as an exponentially damped oscillator.

The  eigenfunctional ${\cal P}_0$, characterizing the long time 
field probability  is then interpreted as the equilibrium 
distribution. Taking $E_0=0$ in Eq.~(\ref{e14}) we  find
\begin{eqnarray}
{\cal P}_{\rm 0}[\pi,\phi]={\cal N} \exp  \left[ - \beta \int d^D x~\left( 
{\pi(x)^2 \over 2} + S(\phi) \right) \right] 
\label{e17}
\end{eqnarray}
with $S(\phi)=\left( {1 \over 2} 
(\nabla \phi(x))^2 + V(\phi) \right)$. Here we took $\Omega=2 \eta/\beta$, 
which is the analog of Einstein's relation for Brownian motion, 
and ensures the balance between fluctuations (sourced by the 
fields $\xi(x,t)$) and dissipation (through the
$- \eta \pi(x,t)$ in Eq.~(\ref{e4b})) at long times.

It is interesting to note that it can be read directly from Eq.~(\ref{e17}) 
that the momentum equilibrium variance
\begin{eqnarray}
\langle \pi(x)^2 \rangle = 1/\beta=T,
\label{e18}
\end{eqnarray}
which expresses equipartition in a relativistic classical field theory.
This quantity is also an excellent thermometer for the dynamical evolution.

Now, because the canonical distribution is Gaussian in $\pi$ we can 
proceed to obtain a reduced distribution in terms of $\phi$ alone.
Performing the Gaussian integral over all $\pi$ we obtain
\begin{eqnarray}
P_{eq}[\phi] = {\cal N}' \exp \left[ - \beta \int d^Dx ~S(\phi) \right],
\label{e19}
\end{eqnarray} 
which is the canonical Boltzmann distribution.   
$P_{eq}[\phi]$ could have been equally obtained
from the static solution of the Fokker-Planck equation associated with the 
perhaps more usual first derivative Langevin equation (see below). 
In this respect we see 
that {\it in equilibrium}  it makes no difference to perform the 
evolution, with or without the second time derivative in Eq.~(\ref{e1}). 
Away from equilibrium, or even if we simply wish to study the fluctuations 
around it, we should take the appropriate form of the evolution, choosing to 
keep or neglect the second time derivative according to the physical picture 
under consideration. The difference between both of these evolutions, 
is expressed in terms of the higher eigenvalues and eigenfunctionals 
associated with the different Hamiltonians, which I discuss 
below.  

For a harmonic potential $V[\phi] = {m^2 \over 2} \phi^2$
we can also read the equilibrium thermal propagator directly from 
Eq.~(\ref{e19}). It is, in momentum space
\begin{eqnarray}
\langle \phi_p \phi_{-p} \rangle = {T \over p^2 + m^2}.
\end{eqnarray}
This is the canonical free Boltzmann propagator, which is the correct 
description for ideal classical waves at finite temperature.

It is illuminating to compare the above framework directly  
with that for the overdamped Langevin evolution. 
Following the same, but somewhat simpler, procedure 
it can be shown \cite{Parisi} that the 
Fokker-Planck Hamiltonian takes the form 
\begin{eqnarray}
{\cal H}_{FP} = - {\delta \over \delta \phi} 
\left[ {\Omega  \over 2 \eta} {\delta \over \delta \phi}  - \nabla^2 \phi
+ {\delta V \over \delta \phi} \right],
\label{e20}
\end{eqnarray}
leading to the eigenvalue functional equation
\begin{eqnarray}
&&- \left\{ {\delta \over \delta \phi} 
\left[ {\Omega  \over 2 \eta} {\delta \over \delta \phi}  - \nabla^2 \phi + 
{\delta V \over \delta \phi} \right] + E_N \right\} P[\phi] = 0, \\
&& \eta \partial_t P[\phi] = - E_N P[\phi]. 
\label{e21}
\end{eqnarray}
Again the canonical distribution is the eigenfunctional $P_0$ corresponding 
to  a zero eigenvalue, i.e.
\begin{eqnarray}
E_0=0; \qquad P_{eq}[\phi] = {\cal N}' e^{- \beta \int d^D x 
{1\over 2} (\nabla \phi)^2  + V(\phi)},
\label{e22}
\end{eqnarray}
where as before the Einstein relation $\Omega= 2 \eta/\beta$ must hold.

\section{Approach to equilibrium}
\label{Sol}

A complete dynamical solution of the ensemble (under average over 
the noise) of fields obeying the Langevin equation (\ref{e4a}-\ref{e4b}) 
can be obtained in 
principle by solving the corresponding Fokker-Planck equation. Unfortunately the task
of solving for these non-linear functional equations in general 
is quite monumental.
The main difficulty is connected with the non-local form of the equations:
in x-space this arises from the Laplacian term, while in k space the 
difficulty is transfered to the non-linear terms.

Below I give closed form solutions to the equations in the 
fields and their conjugate momenta for the particular case of the 
harmonic potential. These are still interesting since 
they offer an alternative to the direct solutions of the Langevin equation,
which, clearly, can also be found in the linear case. In the former picture,
however, the averages over the stochastic fields are already taken into 
account.

It is simpler to build some intuition for the solution of the overdamped 
Fokker-Planck equation first to which I now turn.

\subsection{The overdamped Fokker-Plank equation and its solutions}
\label{sOFP}

The advantage of the harmonic case is that in Fourier space 
different modes decouple in the Langevin equation:
\begin{equation}
\eta \partial_t  \phi_k(t) = (k^2 +m^2) \phi_k(t) + \xi_k (t)
\label{Lang_k}
\end{equation}
where the stochastic fields obey
\begin{eqnarray}
\langle \xi_k(t) \rangle = 0, \quad 
\langle \xi_k(t)\xi_{-k}(t') \rangle = \Omega \delta(t-t')
\end{eqnarray}
and 
\begin{eqnarray}
\phi(x,t) = \int {d^D k \over (2 \pi)^D} \phi_k(t) e^{i k.x} .
\end{eqnarray}
The fact that $\phi(x,t)$ is real implies $\phi_k = \phi_{-k}^*$
and similarly for $\xi_k$. Because of these relations it is more convenient 
to work with $\phi_k^R = {\rm Re}(\phi_k)$ and $\phi_k^I = {\rm Im}(\phi_k)$, which obey 
the Langevin Eq.~(\ref{Lang_k}) with $\xi_k^{R,I}$:
\begin{eqnarray}
 \langle \xi_k(t)^{R,I}\rangle = 0, \quad 
\langle \xi_k^{R,I} (t)\xi_k^{R,I}(t') \rangle = {\Omega \over 2} 
\delta(t-t').
\end{eqnarray}
The Fokker-Planck equation for the modes now is
\begin{eqnarray}
&& \eta \partial_t P_k = -{\cal H}_k P_k \\
&& {\cal H}_k = - {\delta \over \delta \phi_k^{R,I}} 
\left[ {\Omega  \over 4 \eta} {\delta \over \delta \phi_k^{R,I}}  
+ \left(k^2 + m^2 \right) \phi_k^{R,I}  \right].
\label{FPk}
\end{eqnarray}
which holds, as indicated, both for the real and imaginary components
of $\phi_k$. 
The solution of the Fokker-Planck equation then follows as 
\begin{eqnarray}
P = \Pi_{k=0}^\infty \left[ P_k(\phi_k^R) P_k(\phi_k^I) \right]
\end{eqnarray}  
The time independent solution is clearly 
\begin{eqnarray}
P_k(\phi_k^R) P_k(\phi_k^I) \propto \exp \left[ -\beta (k^2 + m^2) \left( 
{\phi_k^R}^2 + {\phi_k^I}^2 \right) \right]
\end{eqnarray}
which leads to 
\begin{eqnarray}
P_0 =&& {\cal N}\exp \left[ -\beta \int_0^\infty  {d^D k \over (2 \pi)^D}~
 (k^2 + m^2) \left( 
{\phi_k^R}^2 + {\phi_k^I}^2 \right) \right]  \\
 =&& {\cal N}\exp \left[ -\beta \int_{-\infty}^{+\infty}
{d^D k \over (2 \pi)^D}~ \phi_k {k^2 + m^2 \over 2} 
\phi_{-k} \right]. \nonumber \\
=&& {\cal N} \exp \left[ -\beta \int_{-\infty}^{\infty} d^D x ~~{1 \over 2}  
\left( \nabla \phi(x) \right)^2 + {m^2 \over 2} \phi^2(x) \right]. \nonumber
\end{eqnarray}

To compute the eigenfunctions corresponding to non-zero eigenvalues
we need to solve the eigenvalue problem
\begin{eqnarray}
\left[ {1 \over 2 \beta} {\delta^2 \over \delta { \phi_k^{R,I}}^2}  
- (k^2 +m^2) \phi_k^{R,I}  {\delta \over \delta \phi_k^{R,I}} 
+E_N \right] F_N = 0
\end{eqnarray} 
where we wrote $P_k = P_0 F_N$.
This equation has a familiar solution. To see this explicitly 
we perform a change of variables to obtain
\begin{eqnarray}
\left[ {\delta^2 \over \delta X }  
- 2 X  {\delta \over \delta X} + {2 E_N  \over k^2+m^2} \right] F_N = 0
\label{Hermite}
\end{eqnarray}
where $X=\sqrt{\beta(k^2 +m^2)} \phi_k^{R,I}$. 
Eq.~(\ref{Hermite}) has the solution
\begin{eqnarray}
&& E_N = N (k^2 +m^2)  \\
&& F_N = H_N (X) = H_N(\sqrt{\beta(k^2 +m^2)} \phi_k^{R,I})
\end{eqnarray}
where $H_N$ is the Hermite polynomial of order $N$. 

From the properties of the Hermite polynomials we know 
that  these solutions are orthogonal under the measure
$e^{-X^2}$, which is just the k-part of the canonical 
distribution. The set of functions $\{H_N(X) e^{-X^2/2} \}$
thus constitutes an orthogonal basis. It can additionally be 
shown trivially that this basis is complete.  

The non-zero Eigenvalues $E_N$ set the time scales for the approach 
to thermal equilibrium, which we will discuss in more detail below.

\subsection{The second order Fokker-Planck equation}
\label{s2FP}

For the second time-derivative harmonic evolution the Langevin
equation becomes
\begin{eqnarray}
&& \partial \phi_k = \pi_k \\
&& \partial \pi_k = - \eta \pi_k - (k^2 +m^2) \phi_k  + \xi_k,
\end{eqnarray}
which as in subsection \ref{sOFP} can be written in terms of real and 
imaginary components. The Fokker-Planck Hamiltonian  for the latter is
\begin{eqnarray}
{\cal H}_k =&& -{\Omega \over 4} {\delta^2 \over \delta {\pi_k^{R,I}}^2}
+ \pi_k^{R,I} { \delta \over \delta \phi_k^{R,I}} \\
&& - {\delta \over \delta \pi_k^{R,I}} \left[ \eta  \pi_k^{R,I} 
+ (k^2 +m^2)  \phi_k^{R,I} \right]. \nonumber
\end{eqnarray}

To find the excited states we proceed as before to factor out the 
canonical distribution $P= F_N P_0$ to get 
\begin{eqnarray}
&& \left\{ -{\Omega \over 4} {\delta^2 \over \delta {\pi_k^{R,I}}^2}
+ \left[ \eta \pi_k^{R,I} -  (k^2 +m^2)\phi_k^{R,I} \right]  
{ \delta \over \delta \phi_k^{R,I}} \right. \nonumber \\
&& \left. \qquad \qquad \qquad 
+ {\delta \over \delta \phi_k^{R,I}}  - E_N \right\} F_N=0.
\label{eigen2}
\end{eqnarray}
We can bring Eq.~(\ref{eigen2}) to the form of Eq.~(\ref{Hermite})
by making the change of variables $X_\pm= a_{\pm} \pi_k^{R,I} + b_{\pm} \phi_k^{R,I}$,
with
\begin{eqnarray}
&&a_\pm^2  = {\beta \over 2} \left[ 1 \pm \sqrt{1 - 4 {k^2 +m^2 \over \eta^2} } \right], \\
&&b_\pm^2 =  {2 \beta \over \eta^2} { (k^2 +m^2)^2 / \left[ 1 \pm \sqrt{1 - 4 {k^2 +m^2 \over \eta^2} 
} \right] }.
\end{eqnarray}
The solutions of Eq.~(\ref{eigen2}) then are
\begin{eqnarray}
&&E_N^\pm = N {a_\pm^2 \eta \over \beta} = N {\eta \over 2} 
\left[ 1 \pm \sqrt{1 - 4 {k^2 +m^2 \over \eta^2} } \right] \label{Solunder} 
\\ &&F_N^\pm = H_N [X_\pm ]. \nonumber
\end{eqnarray} 

These solutions can be brought to products of Hermite polynomials in $\pi_k^{R,I}$ and 
$\phi_k^{R,I}$ alone through the use of the property:
\begin{eqnarray}
H_N[x +y] = 2 ^{-N/2} \sum_{i=0}^N \left( \stackrel{i}{N} \right) H_{N-i} [x \sqrt{2}] H_i [y \sqrt{2}].
\end{eqnarray}  
This property allow us to see that the solutions
\begin{eqnarray}
P_N = \left( F_N [X_+] e^{- E_N^+t}  + F_N [X_-] e^{- E_N^- t} \right) P_0
\label{solcomb}
\end{eqnarray}
form a complete orthogonal basis. 

A few properties of these solutions are  worth pointing out. The coefficients
$a_\pm, b_\pm$ and the eigenvalues $E_N^\pm$ can now be complex numbers. This
happens for $4 (k^2 +m^2) > \eta^2$. In this case $X_+ = X_-^*$, $E_N^+ 
= (E_N^-)^*$ and the combination in (\ref{solcomb}) remains real as required.

The overdamped limit is recovered as $\eta^2 >> k^2 +m^2$. Then
\begin{eqnarray}
&& E_N^- = N {k^2 +m^2 \over \eta}, \label{Over} \\
&& a_-^2= {\beta \over 2} {k^2 + m^2 \over \eta^2} <<  
b_-^2 = {\beta \over 2} (k^2 + m^2).
\end{eqnarray}
The eigenfunctionals associated with $E_N^+ \simeq N \eta 
[1 + (k^2 +m^2)/\eta^2]$ are, in contrast, damped away rapidly.  

The characteristic thermalization times of the system are thus set by 
$t_N = 1/E_N$. For the long wave length modes ($k\simeq0$) we have that
the equilibration time $t_{eq} \sim \eta/(k^2+ m^2)$. 
In the converse limit, of short wavelengths, the decay of all states is 
controlled at leading order  by $t_{eq} \sim 2/ \eta$,  which is  scale independent. 
In this sense the evolution of the short wavelength modes is quantitatively 
different from the overdamped Langevin system.

\section{A mean-field approximation to the non-linear Fokker-Planck dynamics}
\label{MF}

It would no doubt be desirable to be able to solve the Fokker-Planck  
equations for more general (interacting) potentials. At the next level of 
complexity it is possible to attempt a solution of  
the Fokker-Planck equation in a mean-field approximation, which we discuss
in this section.

In order to illustrate the procedure let us consider an interaction  
potential of the form $V_{int}= { \lambda \over 4} \phi(x)^4$.
In momentum space this potential generates a term in the Fokker-Planck 
equation for $\phi_k$ of the form
\begin{eqnarray}
{\delta V_{int} \over \delta \phi_k} = \lambda \int {d^D p \over (2 \pi)^D} 
{d^D q \over (2 \pi)^D}
~\phi_p \phi_q \phi_{k-p-q}.
\end{eqnarray} 
A mean-field approximation to this term can be obtained by taking 
$p=-q$ and the noise average such that
\begin{eqnarray}
{\delta V_{MF} \over \delta \phi_k} = 3 \lambda \left[ \int {d^D p \over (2 \pi)^D} 
\langle \phi_p \phi_{-p} \rangle \right] \phi_{k} \equiv 
3 \lambda G(t) \phi_k.
\end{eqnarray} 
Under this approximation the evolution equation can be written in terms of a 
mean-field Fokker-Planck operator where $V_{MF}$ substitutes the potential $V$, with
\begin{eqnarray}
V_{MF} = V_{0} +  {3 \lambda \over 4} G(t)^2,
\end{eqnarray}
where $V_{0}$ corresponds to the purely harmonic case, 
together with the self-consistency condition 
\begin{eqnarray}
G(t) = \int {d^D p \over (2 \pi)^D}  \int d \phi_p~ \phi_p \phi_{-p} P[\pi_q,\phi_p,t].
\end{eqnarray}
It is then clear that the under this Gaussian approximation the theory 
remains harmonic, but with a time-dependent (self-consistently determined) mass.

Unfortunately the time dependence of the mean field potential 
destroys the separability of the solution into a function of time, 
and a functional 
of the fields and their conjugate momenta. Closed form solutions are thus 
difficult to construct, which is of course also the case for the 
Langevin equation under a similar approximation.
Numerical solutions of the this mean field Fokker Planck equation can 
nevertheless be easily obtained. 

As a starting point let us again consider the overdamped case first. 
Mean field approximations correspond in general to Gaussian probability 
distributions. In the overdamped case, this can only lead to 
\begin{eqnarray}
P_{MF} [\phi_k,t] = {\cal N}_k(t) \exp{\left[ 
- \phi_{-k} {A_k^2(t) \over 2} \phi_k \right]},
\end{eqnarray} 
where the normalization ${\cal N}(t)$ is a function of $A_k^2(t)$
\begin{eqnarray}
{\cal N}_k(t) = A_k(t)/\sqrt{\pi},
\end{eqnarray}
and must therefore be time-dependent. This does not of course introduce
any additional dynamical freedom.

The mean-field Fokker-Planck equation for $P_{MF} [\phi_k,t]$ translates 
into an equation for the coefficients $A_k(t)$ of the form
\begin{eqnarray}
&& \eta {d A_k \over dt} = \left[ k^2 + m^2 
+ 3 \lambda G(t) -A_k^2 T \right] A_k, \label{MF1} \\
&& G(t) = \int {d^D p\over (2 \pi)^D} {1 \over A_p^2}. \label{MF2}
\end{eqnarray}
Clearly solving the Fokker-Planck equation in this mean-field approximation
requires the solution of a system of coupled  ordinary differential 
equations, one for every mode $k$. This is similar in effort to the solution
of the Langevin mean-field equation with the advantage that the stochastic 
average is already taken into account. This latter step corresponds to the 
functional integral necessary to compute $G(t)$ in the Fokker-Planck picture, 
which can be given in closed form because $P_{MF}$ is Gaussian.

The canonical (mean-field) distribution is of course a static solution of Eqs~(\ref{MF1}-\ref{MF2}) 
where $A_k$ obeys
\begin{eqnarray}
A_k^2 = {k^2 + m^2 \over T} + 3 {\lambda \over T}  \int {d^D p\over (2 \pi)^D} {1 \over A_p^2},
\label{GapMF}
\end{eqnarray}
It is then clear that for vanishing interactions $\lambda=0$, the free Boltzmann propagator 
is the solution of Eq.~(\ref{GapMF}). 
In general because the mean-field contribution is momentum independent 
Eq.~(\ref{GapMF}) are equivalent to the 'gap' equation
\begin{eqnarray}
A_k^2 &&= {k^2 + m^2 + \Delta m^2  \over T}, \\
\Delta m^2 &&= 3 \lambda \int {d^D p\over (2 \pi)^D} {T \over k^2 +m^2 + \Delta m^2}.
\end{eqnarray}

For the second order Fokker-Planck equation the most general Gaussian 
function of the field and momentum modes is of the form
\begin{eqnarray}
&&P_{MF} [\pi_k,\phi_k,t] = \\
&& {\cal N}_k(t)\exp \left\{ - \left[ {A^2_k \over 2} 
\phi_k \phi_{-k} + {B^2_k \over 2} \pi_k \pi_{-k} + C^2_k {\rm Re}(\pi_k \phi_{-k}) \right] \right\}.
\nonumber 
\end{eqnarray}
The normalization is now a function of all $A^2_k,B^2_k,C^2_k$
\begin{eqnarray}
{\cal N}_k = {1 \over \pi}\sqrt{A^2_k B^2_k - C^4_k}. 
\end{eqnarray} 
The equations of motion for $A^2_k,B^2_k$ and $C^2_k$ then follow from
the second order Fokker-Planck equation:
\begin{eqnarray}
&& {d A_k \over dt} =  {C_k \over A_k} \left[ k^2 + m^2 + 3 \lambda G(t) - \eta T C_k^2 \right] \\
&& {d B_k \over dt} = \eta B_k \left[ 1 - T B_k^2 \right]  \\
&& {d C_k \over dt} = {\eta C_k \over 2} \left[ 1 - 2 T B_k^2 \right] \nonumber \\
&& \qquad \qquad -{1\over 2 C_k} \left[A_k^2 -  \left(
k^2 + m^2 + 3 \lambda G(t) \right) B_k^2 \right],
\end{eqnarray}
where
\begin{eqnarray}
G(t) = \int {d^D p \over (2 \pi )^D} {B_p^2 \over A_p^2 B_p^2 - C_p^4}.
\end{eqnarray}
This system has as static solutions the mean-field canonical distribution 
namely
\begin{eqnarray}
&&A_k^2 = {k^2 +m^2 +\Delta m^2 (T) \over T}, \\
&&B_k^2 = 1/T = \beta \\
&&C_k^2 = 0,
\end{eqnarray}
where the equation for $A_k$ is simply (\ref{GapMF}). 
Other static solutions exist though. Additional stationary 
distributions distinct from thermal equilibrium also occur in the 
microcanonical evolution of the class of models considered here 
\cite{BW} and seemingly result  from the truncation
of the system at any finite order in a hierarchy of correlators.    
The overdamped limit, $k^2 + m^2 + \Delta m^2(t) << \eta$ 
is obtained when $C_k^2 \simeq A_k^2/\eta << B_k^2, A_k^2$.

Again we see that the effort in solving the mean-field second order 
Fokker-Plank equation is comparable to its Langevin counterpart, but with 
the stochastic averages taken into account. A renormalization scheme for
rendering these equations ultraviolet cutoff independent 
can be constructed, if desired, following standard techniques for 
mean-field evolutions \cite{Cooper}.

\section{The uses of the closed form solutions of the Fokker-Planck equation}
\label{Dyn}

The solutions presented above permit some analysis of several 
important dynamical situations. 

In many cases in field theory the harmonic solutions can reveal the essential 
physics of the (short time) evolution. This is true for example
for the motion of soft fields in the vicinity of a second order  phase transition, 
where only the lowest relevant operators are needed,
and for some situations when (spinodal) instabilities can develop, such as 
in the early stages of reheating after cosmological inflation or in the 
aftermath of a violent pressure quench, in some condensed matter experiments. 

In this section I analyze these two situations in the context of 
the solutions  of section \ref{Sol}.

\subsection{Critical Dynamics of the fields and the theory of 
topological defect formation at second order transitions}

In the vicinity of a symmetry breaking second order 
transition the physical mass squared of the fields will approach 
zero, with a universal critical exponent $\nu$,  
\begin{eqnarray}
m^2(T) \sim m^2 \vert {T - T_c \over T_c} \vert^\nu, \nonumber
\end{eqnarray}
where $T_c$ is the critical temperature. This can be seen already 
at first order in perturbation theory, albeit only with the 
mean-field value of $\nu=1/2$. In 3D we have
\begin{eqnarray}
m^2(T) \sim  -m^2 + 3 \lambda \int^\Lambda {d^3 k \over (2 \pi)^3} {T \over k^2  } = 
-m^2 + {3 \lambda \over 2 \pi^2} \Lambda T.
\label{tadpole}
\end{eqnarray}
$\Lambda$ is the ultraviolet cutoff which in many cases has physical meaning
because generally scalar field theories are effective low energy models.
An example is the scale of separation between the true quantum behavior 
of excitations at large $k^2$ and the effective classical dynamics for the 
long wave-length modes which we describe
stochastically. In this case, at high 
temperature and small coupling $\lambda$, $\Lambda \simeq  T$.

The interesting fact about $m^2(T) \rightarrow 0$ 
at criticality is that it supplies us with 
an arbitrarily long time scale separation between the 
thermalization time for the short and long 
wave-length modes. As we discussed in section \ref{Sol} 
the long wave length modes $k^2 \simeq 0$ thermalize in a typical  time-scale
\begin{eqnarray}
t_{eq} \simeq {\eta \over m^2(T)} {\rightarrow}_{T\rightarrow T_c}  + \infty,
\label{critslow}
\end{eqnarray} 
which is the expression of critical slowing down in our stochastic system. On the other 
hand the short-wave length modes thermalize instead on a characteristic time scale
\begin{eqnarray}
t_{eq} \simeq 2/\eta.
\end{eqnarray}
It is worth commenting of the form (\ref{critslow}). It shows that even in the 
context of a second order 
in time Langevin evolution the long wavelength modes are effectively 
overdamped in the critical domain. This is the essence of the perhaps more familiar 
TDGL evolution, which is an expansion in the lowest number of relevant field operators 
that can only be justified rigorously in the critical domain of a second order transition. 
We see therefore that our Langevin description encapsulates the TDGL dynamics 
but in addition also applies more generally.    

Now, particularly in $D=3$ (as compared to lower dimensions) 
the temperature corrections to the mass are dominated by the 
small wave-length modes. Imagine then that the bath temperature is changed to a new value 
$T_f$ in the vicinity of the phase transition. Over a time $t \sim 2/\eta$ 
the thermal mass will then adapt to its new small value $m^2(T_f)$. 
In contrast the long wave-length modes require a much 
longer time to rethermalize and stay away from thermal equilibrium 
for a time $t\sim \eta/m^2(T_f) >> \eta$. 

This imbalance is at the heart of our current  understanding of 
the {\it dynamics} of second order
phase transitions and constitutes in particular the essence of the 
Kibble-Zurek theory of topological defect 
formation. The familiar response time $\tau(T)$ is here given 
by Eq.~(\ref{critslow}). Incidentally the 
temperature dependence of (\ref{tadpole}) also implies that 
whatever the time behavior of the external bath temperature 
the dependence of $m^2(T)$ on time will be the same, i.e. for example for 
a linear external drive of the bath temperature one also obtains  $m^2(t)\propto t $, 
again on timescales $t \stackrel{>}{\sim} 2/\eta$.  

The consequences of this imbalance are very important. 
By crossing a second order phase transition 
at some finite rate $\tau_Q$ imposed externally there will be a time
in the vicinity $T_c$ when the long-wave length modes fall out of 
thermal equilibrium. Because the system always rethermalizes in its 
small scales first, soft long wavelength non-trivial field 
configurations can persist for a long time. In particular topological defects 
can be ``formed'' in this way.

\subsection{Spinodal instabilities}

Another interesting application  of the harmonic solutions 
of section \ref{Sol} is the onset of (spinodal) instabilities.
This situation assumes an initial instability characterized {\it eg.} by  
a coherent ({\it i.e.} quasi-spatially homogeneous) field, close to the 
origin $\phi=0$ of a double-well potential. In addition there may be 
fluctuations (thermal and/or quantum) about this mean field but these 
must be tuned to be small and are often taken to be Gaussian and white. 

In these circumstances $m^2$ is initially 
negative and the long wave-length modes $k^2 < -m^2$ will 
develop instabilities characterized by exponential growth of their 
amplitude at early times. In the limit of vanishing stochastic fields 
this behavior can be captured by solving the linearized Langevin 
equation \cite{Grant,Jacek,Ray}, whose solutions are of course simple 
exponentials of $\pm i \sqrt{k^2 -m^2}t $. Linear dependences of 
$m^2$ on time can also be captured in similar ways.

The advantage of using the set of solutions (\ref{Solunder}) in this context 
is that the noise effects are then better taken into account. 
For the linearized problem the set of eigenfunctions and eigenvalues 
given in (\ref{Solunder}) is exact. The behavior of these solutions 
in the overdamped case, where the eigenvalues are 
\begin{eqnarray} 
E_N = N { \left( k^2 -m^2 \right) \over \eta},
\label{Eover}
\end{eqnarray} 
is obvious: whenever $k^2 -m^2 < 0$ all excited states display exponential 
growth. It is worth noting that it is the highest 
lying states (i.e. those with largest $N$) that grow the fastest,
leading to spectacularly out of equilibrium configurations, 
at least relative to the initial state. The short wavelength modes 
are of course still dissipated away on a time scale  $t \sim 2/\eta$ 
so these excited states are highly biased towards the infrared as 
is well known from direct numerical studies.

For the  second order Fokker-Planck equation the eigenvalues 
are instead (\ref{Solunder}). 
It is then clear that there will be growing modes as 
long as  $k^2 < -m^2$. In these circumstances, for small
but negative $m^2/\eta$, the eigenvalues coincide  with (\ref{Eover}). 
For the second time derivative evolution however there 
will be both exponentially growing and decaying modes.

The validity of the linear approximation breaks down when the 2-point function
$G(t)$, which controls the leading corrections to the mass squared, becomes  
of order the negative bare mass squared. In the Fokker-Planck picture 
this is an average which correctly accounts for the noise effects on the short 
wavelength modes, which are quasi-Gaussian (perturbative) 
in any case.  The value of $G$ in linearized evolution
can clearly be computed in closed form since all integrations are over 
polynomials of the  fields times a Gaussian. 

Computations of eg. $G(x,t)$ at the point where the linearized 
approximation fails have been shown to be useful \cite{Grant}, {\it eg}. 
in the computation of topological defect formation through 
the insertion of $G(x,t)$ into the well-known Halperin formula 
for counting field zeros, provided there is little energy left in the 
system at this point to cause substantial subsequent transport towards 
short wavelengths, (see \cite{AB} for counterexamples).

\section{Discussion and Conclusions}
\label{Conc}  

In this paper I discussed general properties of the second order  
Langevin scalar field evolutions as well as of some of its related models. 
The associated Fokker-Planck equations were solved in closed
form for the harmonic potential and a mean-field approximation 
was devised for both second order and overdamped cases. 
The analogous treatment for the 
stochastic nonlinear Schr\"odinger equation was given in Appendix A. 
The latter can be solved numerically with an effort that is comparable
to  the mean-field Langevin equation, but has the advantage of 
already including the effects of the averages over the stochastic fields.

Although limited to special cases closed form solutions of the Fokker-Planck 
equations, have been shown to be very useful in providing us with analytical qualitative 
(and sometimes quantitative) insights into the evolution of the fully 
non-linear  field theoretical system, especially in determining time scales 
for thermalization of several parts of the system. 
Through this analysis I showed 
how fundamental ingredients of the dynamics of second order phase transitions
are embodied in these effective models and how they generalize the well known
TDGL evolution in the critical domain.

In particular in the vicinity of $T_c$, 
where $m^2(T)\rightarrow 0 $, one obtains true time scale separation, with 
the small wavelength field modes thermalizing in a time 
$t_{eq} \sim \eta^-1$, while the long wave length modes take a time 
$t_{eq} \sim \eta/m^2(T) >> 1/\eta$. In these circumstances, over times 
$1/\eta << t << \eta/m^2(T)$, 
one can take the short wavelength modes to be thermalized 
at the temperature of the stochastic thermal bath, and include their effects
on the long-wavelength modes through the temperature dependence of the parameters
in their potential. This treatment allows us to discuss in qualitative correct terms 
the evolution  of the long wavelength modes in the critical domain, the 
phenomenon of critical slowing down and the ingredients of the 
theory of topological defect formation in these models.

\section{Acknowledgments}

It is a pleasure to thank  N.~Antunes, S. Habib, G.~Lythe and 
W. H. Zurek for useful comments. 
This research was supported by the U.S. Department 
of Energy, under contract W-7405-ENG-36.

\section*{Apendix A}

In this appendix I repeat the analysis applied to the Langevin 
equations~(\ref{e1})
to the case of a stochastic non-linear Schr\"odinger equation (sNLS), 
known in some instances also as the stochastic Gross-Pitaevsii equation.
This model is important in the study of non-relativistic many body systems. 
Many examples exist that are thought to obey this effective dynamics, ranging 
from strong type II (or hard) superconductors and superfluid $^4$He 
to atomic Bose-Einstein condensates and light propagation 
in non-linear media.    

The sNLS is of the form
\begin{eqnarray}
-\left(i+\eta \right) \partial_t \psi = - \nabla^2 \psi + {\delta V \over 
\delta \psi^*} +\xi(x,t),
\label{sNLS}
\end{eqnarray}  
where $\psi= \psi(x,t)$ is a complex field in space and time. 
The potential $V$ is usually taken to be
\begin{eqnarray}
V \left[ \vert \psi \vert^2 \right] = 
m^2  \vert \psi(x,t) \vert^2 + 
{\lambda \over 2}  \vert \psi(x,t) \vert^4,
\end{eqnarray}
where $m^2$ can be positive or negative. This theory is essentially the 
non-relativistic  version of the second order in time Langevin system for a
complex scalar field. The theory is manifestly invariant 
under global phase $U(1)$ transformations.
Its Hamiltonian is 
\begin{eqnarray}
H = \int d^Dx \left\{ \vert \nabla \psi \vert^2 
+ V\left[\vert \psi \vert^2 \right] \right\}.
\end{eqnarray}

Although the stochastic terms in Eq.~(\ref{sNLS}) may be representative 
of additional intrinsic degrees of freedom, 
it may also be possible to actually build 
an experimental situation in which the many-body system is driven 
externally and has losses to the outside so as to realize (\ref{sNLS}). 
In this latter picture the requirements are that the driving field should 
be phase incoherent (at least over some short characteristic time and 
spatial scales) and that the losses would be proportional to the 
frequency of the excitations in the system.  
Perhaps the most challenging feature  would be to find 
a trap or mirror whose transmission is linear in the range of 
frequencies expected in the system.

The Fokker-Planck equation corresponding to this evolution is
\begin{eqnarray}
\partial_t P =&& \left\{ {1 \over i +\eta} {\delta \over \delta \psi}
\left[ 2 {\delta H \over \delta \psi^*} 
+ {\Omega \over -i +\eta} {\delta \over \delta \psi^*} \right] 
\right. \\
&& \left. \qquad \qquad + {1 \over -i +\eta} {\delta \over \delta \psi^*} 
\left[ 2 {\delta H \over \delta \psi} 
+ {\Omega \over i +\eta} {\delta \over \delta \psi} \right] \right\} P 
\nonumber
\end{eqnarray}
where $\Omega = 2 \eta T$ as before.
The system thermalizes to its canonical distribution 
\begin{eqnarray}
P_0 = {\cal N} \int D \psi ~D \psi^* \exp \left[ - \beta H \right].
\end{eqnarray}

The set of eigenvalues and eigenvectors for the harmonic problem are
\begin{eqnarray}
&& E_N = 2 N \eta { k^2 + m^2 \over 1 + \eta^2}, \\
&& F_N = H_N \left[ \sqrt{\beta (k^2 +m^2) \psi_k \psi_k^*} \right].
\end{eqnarray}
The factor of 2 in $E_N$ relative to (\ref{Over}) accounts for 
2 real fields in the complex quantity $\psi$, instead of one. 
Thus, the equilibration time scale is
\begin{eqnarray}
t_{eq} \sim {1 +\eta^2 \over 2 \eta \left(k^2 +m^2 \right) },
\end{eqnarray}
which again displays critical slowing down for 
$k^2 \simeq 0$ and $m^2(T) \rightarrow 0$.

It is clear that the overdamped limit is recovered 
for large $\eta$, as expected.
We therefore see that in the overdamped limit all three models considered 
in this paper lead to the same characteristic time scale for equilibration
as could be expected on general grounds. The differences arise for the short 
wavelength modes in the system and/or for small $\eta$.

An experimental realization of (\ref{sNLS}) would allow for the driving of 
the physical system across its phase 
transition and more generally to canonical equilibrium at an arbitrary 
temperature, by changing the intensity of the phase incoherent 
driving field $\xi$. 
If $\xi$ had a phase coherent component, in  addition to the incoherent 
piece, the drive would make the system behave like an XY magnet at finite 
temperature in the presence of a magnetic field in the XY plane. 
This would allow for experiments in non-linear optics (where this effect is known
as phase pinning)  or atomic Bose-Einstein condensates to explore regimes similar 
to those realized in {\it eg.} high-Tc superconductors.

\end{document}